\documentclass[aps,prc,twocolumn,showpacs,floatfix,nobibnotes,nofootinbib,preprintnumbers,superscriptaddress,amsmath,amssymb]{revtex4-1}

\usepackage{graphicx}
\usepackage{epsfig}
\usepackage{bm}
\usepackage{color}
\usepackage{float}
\usepackage{dcolumn}
\usepackage{multirow} 
\usepackage{hyperref}
\usepackage{algorithmicx}
\usepackage{algpseudocode}

\graphicspath{{.}{../}}

\newcommand{\beq}{\begin{equation}}
\newcommand{\eeq}{\end{equation}}
\newcommand{\beqa}{\begin{eqnarray}}
\newcommand{\eeqa}{\end{eqnarray}}

\begin{document}
\title{Open $sd$-shell nuclei from first principles}
\thanks{This manuscript has been authored by UT-Battelle, LLC under
  Contract No. DE-AC05-00OR22725 with the U.S. Department of
  Energy. The United States Government retains and the publisher, by
  accepting the article for publication, acknowledges that the United
  States Government retains a non-exclusive, paid-up, irrevocable,
  world-wide license to publish or reproduce the published form of
  this manuscript, or allow others to do so, for United States
  Government purposes. The Department of Energy will provide public
  access to these results of federally sponsored research in
  accordance with the DOE Public Access
  Plan. (http://energy.gov/downloads/doe-public-access-plan).}

\author{G.~R.~Jansen}
\affiliation{National Center for Computational Sciences, Oak Ridge National
Laboratory, Oak Ridge, TN 37831, USA}
\affiliation{Physics Division, Oak Ridge National
Laboratory, Oak Ridge, TN 37831, USA}

\author{M.~D.~Schuster}
\affiliation{National Center for Computational Sciences, Oak Ridge National
Laboratory, Oak Ridge, TN 37831, USA}

\author{A.~Signoracci}
\affiliation{Physics Division, Oak Ridge National
Laboratory, Oak Ridge, TN 37831, USA}
\affiliation{Department of Physics and Astronomy, University of Tennessee,
Knoxville, TN 37996, USA}

\author{G.~Hagen}
\affiliation{Physics Division, Oak Ridge National Laboratory,
Oak Ridge, TN 37831, USA} 
\affiliation{Department of Physics and Astronomy, University of Tennessee,
Knoxville, TN 37996, USA} 

\author{P.~Navr\'{a}til}
\affiliation{TRIUMF, 4004 Wesbrook Mall, Vancouver,
British Columbia, V6T 2A3 Canada}

\begin{abstract}
  We extend the {\it ab initio} coupled-cluster effective interaction
  (CCEI) method to open-shell nuclei with protons and neutrons in the
  valence space, and compute binding energies and excited states of
  isotopes of neon and magnesium. We employ a nucleon-nucleon and
  three-nucleon interaction from chiral effective field theory evolved
  to a lower cutoff via a similarity renormalization group
  transformation. We find good agreement with experiment for binding
  energies and spectra, while charge radii of neon isotopes are
  underestimated. For the deformed nuclei $^{20}$Ne and $^{24}$Mg we
  reproduce rotational bands and electric quadrupole transitions
  within uncertainties estimated from an effective field theory for
  deformed nuclei, thereby demonstrating that collective phenomena in
  $sd$-shell nuclei emerge from complex {\it ab initio} calculations.
\end{abstract}

\maketitle 

{\it Introduction} -- Nuclei are complex many-body systems that
present us with a wealth of interesting quantum mechanical phenomena
that emerge along the entire chart of nuclei. These phenomena involve:
exotic clustering behavior and extended density distributions of
loosely bound nuclei \cite{epelbaum2011,tanihata2013}, melting and
re-organization of shell-structure in neutron nuclei
\cite{hagen2012b,holt2012,steppenbeck2013}, Borromean nuclei
\cite{tanihata1985,zhukov1993}, and emergence of collective behavior
in nuclei, such as rotational and vibrational states
\cite{bohr1952,bohr1953} as well as nuclear super conductivity and pairing
\cite{bohr1958}.

Recently there has been an explosion of nuclear many-body methods with
a sufficiently soft computational scaling to allow a reliable
description of binding energies and spectra in nuclei up through the
$sd$-shell starting from nucleon-nucleon and three-nucleon forces from
chiral effective field theory (EFT)
\cite{hebeler2015,simonis2015,binder2013b,hergert2014,ekstrom2015,hagen2015}. In
spite of this progress, emergence of collective phenomena in nuclei
still poses significant challenges to \emph{ab initio} methods.
Rotational states in $p$-shell nuclei have been successfully computed
in the no-core shell-model and in Green's-function Monte-Carlo
approaches
\cite{caurier2001,pieper2004,caprio2015,capriotemp,maris2015}, while
in the $sd$-shell, deformed nuclei have only been accurately described
in shell-model calculations using phenemenological
interactions~\cite{caurier2005}.  A symplectic approach has been
proposed~\cite{dytrych2013} to enable extension of the no-core
shell-model to larger model spaces and higher-mass nuclei, yet
prototypical deformed nuclei like $^{20}$Ne and $^{48}$Cr remain out
of reach in the aforementioned approaches.  Furthermore, as deformed
nuclei are truly open-shell, they are inaccessible via the typical
implementation of many-body expansion methods like coupled cluster
theory \cite{hagen2014}, self-consistent Green's function methods
\cite{dickhoff2004}, and in-medium similarity renormalization group
(SRG) methods \cite{hergert2013b}, that normally rely on a spherical,
closed-shell mean field.  Extensions to open-shell nuclei via
Bogoliubov coupled cluster theory \cite{signoracci2015},
Gorkov-Green's function methods ~\cite{soma2013}, and multi-reference
in-medium similarity renormalization group methods \cite{hergert2014}
are underway, but are limited to semi magic nuclei thus far.

Recently, non-perturbative effective interactions for the shell-model
were derived from coupled-cluster theory and in-medium SRG
starting from chiral nucleon-nucleon and three-nucleon forces
\cite{bogner2014, jansen2014}.  The CCEI approach is based on the
valence-cluster expansion of the Hamiltonian that was first proposed
within the framework of the no-core shell-model~\cite{barrett2013},
and applied to $p$-shell
nuclei~\cite{navratil1997,lisetskiy2008}. More recently the no-core
shell-model was used to construct non-perturbative shell-model
interactions for light fluorine isotopes based on nucleon-nucleon
interactions only~\cite{dikmen2015}.

In this work we extend the CCEI method to deformed nuclei in the
$sd$-shell with both protons and neutrons in the valence
space. Furthermore, we show that other observables such as charge
radii can be consistently computed within CCEI. Diagonalization of the
resulting interactions yields the fully-correlated wavefunction in the
reduced model space, which is sufficient to describe the properties of
deformed nuclei \cite{brown2006}. This enables the description and
identification of rotational bands, which we compare with predictions
from an EFT developed for deformed
nuclei~\cite{papenbrock2011,papenbrock2014,coelloperez2015}.

{\it Theory} -- To minimize spurious center-of-mass motion
\cite{hagen2009a}, our coupled-cluster calculations start from the
intrinsic, $A$-dependent Hamiltonian,
\begin{equation}
  \label{intham}
  \hat{H} = \sum_{i<j}\left({({\bf p}_i-{\bf p}_j)^2\over 2mA} + \hat{V}
    _{NN}^{(i,j)}\right) + \sum_{ i<j<k}\hat{V}_{\rm 3N}^{(i,j,k)}.
\end{equation}
We utilize the same Hamiltonian as in our prior
work,~\cite{jansen2014}, with an initial next-to-next-to-next-to
leading order two-body chiral interaction and a next-to-next-to
leading order three-body chiral interaction, evolved consistently to a
lower momentum scale via a similarity renormalization group
transformation. This interaction has been demonstrated to yield
accurate binding energies and spectra in and around isotopes of oxygen
\cite{hergert2013,cipollone2013,bogner2014,jansen2014}. Here we
explore how this Hamiltonian performs in isotopes
of neon and magnesium. We perform coupled-cluster calculations in
thirteen major oscillator shells with the frequency $\hbar \omega =20$
MeV, and utilize the normal-ordered two-body approximation for the
three-nucleon force with the additional three-body energy cut
$E_{3\text{max}}=N_1+N_2+N_3 \leq 12$, where $N_i$ refers to the major
oscillator shell of the $i^{th}$ particle (see Ref.~\cite{jansen2014}
for more details). In addition, we derive the coupled-cluster
effective interactions (CCEI) based on a core of mass $A_c$, by
expanding the Hamiltonian Eq.~\eqref{intham} in a form suitable for
the shell model (i.e. the valence-cluster expansion
\cite{lisetskiy2008}),
\begin{equation}
  \label{ham}
  H_{\rm CCEI} = H_0^{A_c} + H_1^{A_c+1} + H_2^{A_c+2} + \ldots,
\end{equation}
i.e. for the core, one-body, two-body, and higher body valence-cluster
Hamiltonians. As pointed out in Ref.~\cite{dikmen2015} the valence
-cluster expansion is not uniquely defined as one can either choose to
set the mass $A$ that appears in intrinsic kinetic energy of the
individual parts equal to the mass $A$ of the target nucleus, or set
it equal to the mass $A$ of the core, one-body, two-body, and higher
body parts. Both choices will reproduce the exact result in the limit
of including all terms in the valence-cluster expansion. While in
Ref.~\cite{jansen2014} we made the former choice, in this work we
choose the latter as defined by Eq.~\eqref{ham} and truncate the
cluster expansion at the two-body level. This choice tremendously
simplifies the calculations since we can use the same effective
shell-model interaction for all nuclei, furthermore this choice
guarantees that the Hamiltonians for $A_c, A_{c+1}$, and $A_{c+2}$ are
translationally invariant (see Supplemental Material for a
quantitative comparison between these two choices in computing excited
states and binding energies in oxygen isotopes). In this work we
compute the ground state of the $A_c$ nucleus using the
coupled-cluster method in the singles-and-doubles approximation with
the $\Lambda$-triples correction treated perturbatively
($\Lambda$-CCSD(T)) \cite{taube2008,hagen2010b}, while the one- and
two-particle-attached equation-of-motion coupled-cluster (EOM-CC)
methods are used to compute the ground and excited states of the
$A_c+1$ and $A_c+2$ nuclei
\cite{gour2006,jansen2011,jansen2012,shen2014}. In this work we define
our valence space by the $sd$-shell, and we use the Okubo-Lee-Suzuki
similarity transformation \cite{okubo1954,suzuki1982,kvaal2008} to
project the one- and two-particle-attached EOM-CC eigenstates with the
largest overlap with the model space onto two-body valence-space
states. From the non-Hermitian coupled-cluster procedure, one can
obtain a Hermitian effective Hamiltonian for use in standard shell
model codes by constructing the metric operator $S^\dagger S$ where
$S$ is a matrix that diagonalizes $ {H}^A_{\rm CCEI} $; the Hermitian
shell-model Hamiltonian is then $\left[S^\dagger S\right]^{1/2}
{H}^A_{\rm CCEI} \left[S^\dagger S\right]^{-1/2}$
\cite{scholtz1992,navratil1996}.

Any operator $O$ can be expanded in a similar form to Eq.~\eqref{ham},
suitable for the shell model (see
Refs.~\cite{navratil1997,lisetskiy2009} for details), $O_{\rm CCEI} =
O_0^{A_c} + O_1^{A_c+1} + O_2^{A_c+2} + \ldots$, with a consistent
(i.e. identical to that of the Hamiltonian) Okubo-Lee-Suzuki
transformation and metric operator. In this way, any operator which
can be computed for all many-body states in the $A_c,A_c+1,A_c+2$
systems in coupled-cluster theory can be used to define a
valence-space operator.

{\it Results} -- As we in this work adopt a different definition for
the valence-cluster expansion as used in Ref.~\cite{jansen2014}, we
would first like to address the accuracy of our calculations. Again, we find
good agreement betwen full-space CC and CCEI for binding energies
in the oxygen isotopes, and for low-lying excited states in
$^{22,24}$O. In particular, for $^{22}$O CCEI yields a $J^\pi = 2^+$
excited state at 2.6~MeV and a $J^\pi = 3^+$ excited state at 3.8~MeV,
while full-space EOM-CC with singles and doubles excitations
(EOM-CCSD) yields the corresponding excited states at 2.5~MeV and
3.8~MeV, respectively. For the $J^\pi = 2^+ $ and $J^\pi = 1^+$
excited states in $^{24}$O CCEI gives 5.7~MeV and 6.4~MeV, while
EOM-CCSD yields 6.0~MeV and 6.4~MeV, respectively. The CCEI results
for the binding energies of $^{22,24}$O are 162.2~MeV and 168.1~MeV,
while the corresponding full-space $\Lambda$-CCSD(T) results are
$162.0$~MeV and $170.2$~MeV, respectively. We refer the reader to the
Supplement Material for a more detailed comparison between
full-space coupled-cluster calculations and results obtained using the
$A$-dependent and $A$-independent choices in the CCEI method.

\begin{figure}[tbh]
  \includegraphics[width=1.0\columnwidth]{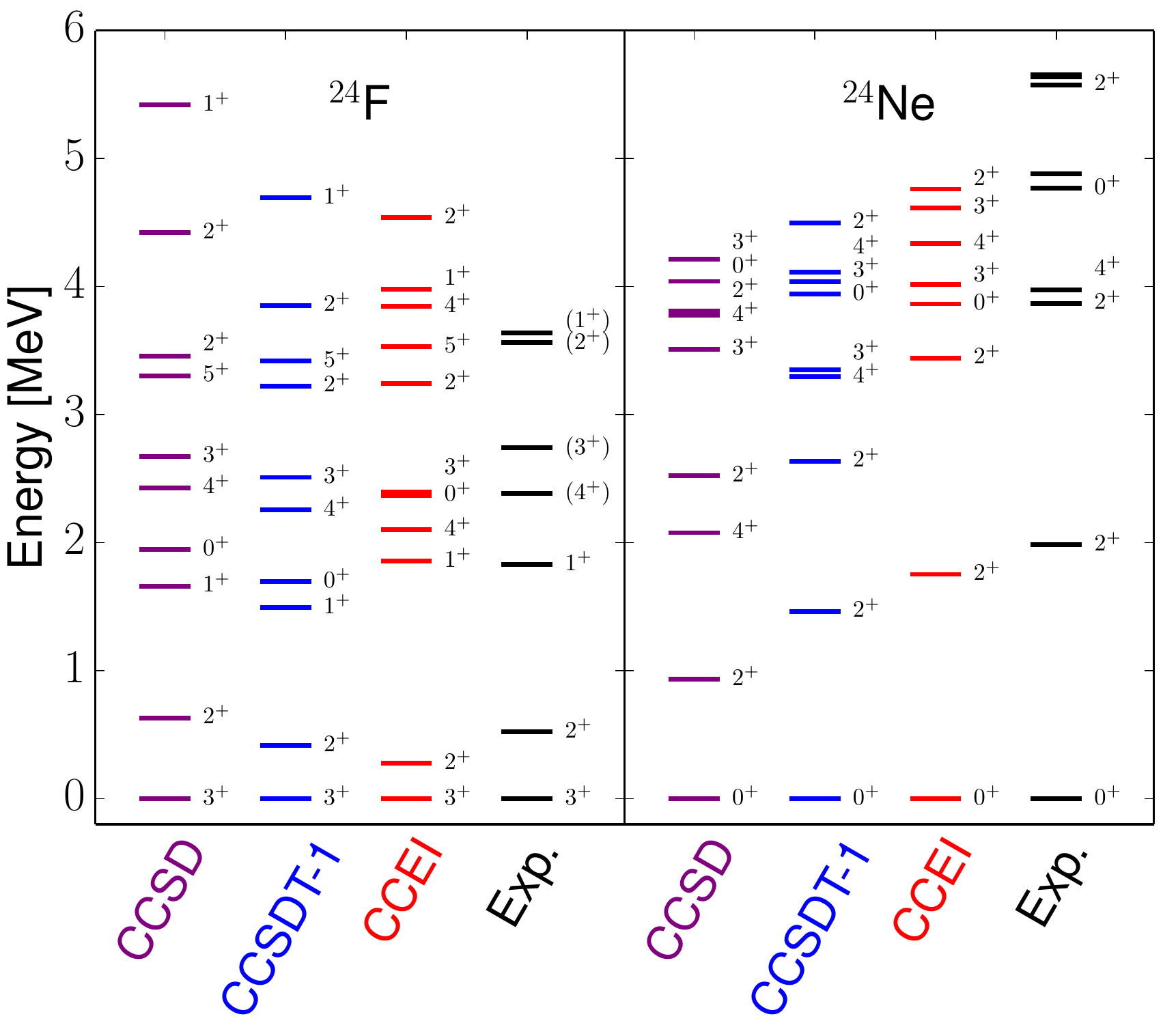}
  \caption{(Color online) Excited states of $^{24}$F (left panel) and
    $^{24}$Ne (right panel) computed from the charge-exchange
    EOM-CCSD, EOM-CCSDT-1, CCEI, and compared to data.}
  \label{FNe_benchmark}
\end{figure}

In this work we extend the CCEI approach to nuclei with protons and
neutrons in the valence space, and to gauge the accuracy of CCEI for
these systems we benchmark against the full-space charge-exchange
EOM-CC method~\cite{ekstrom2014} for ground- and
excited states in $^{24}$F and $^{24}$Ne. To obtain a more
precise calculation of $^{24}$Ne as a double charge-exchange
excitation from the ground-state of $^{24}$O, we extend the
charge-exchange EOM-CC method beyond the two-particle-two-hole
excitation level (EOM-CCSD), and include the leading-order
three-particle-three-hole ($3p$-$3h$) excitations defined by the
EOM-CCSDT-1 method \cite{watts1995}. Since this approach is rather
costly in terms of computional cycles and memory, we introduce an
active-space~\cite{gour2005} truncation on the allowed $3p$-$3h$ excitations in
the unoccupied space defined by an energy cut $e_{3\text{max}}=N_1+N_2+N_3$
(similar to the $E_{3\text{max}}$ cut of the three-nucleon force). This
approach allows us to compute ground- and excited states of nuclei
that differ by two units of the z-component of the total isospin
(double charge-exchange) from the closed (sub-)shell reference
nucleus, and here we present the first application of this method to
ground and excited states of $^{24}$Ne. Figure~\ref{FNe_benchmark}
shows the low-lying spectra of $^{24}$F and $^{24}$Ne computed with
full-space (double) charge-exchange EOM-CCSD and EOM-CCSDT-1 and CCEI,
including a comparison to data.  The agreement between the CCEI and
full-space charge-exchange EOM-CCSDT-1 for $^{24}$F is overall good,
and we see that the effect of including $3p$-$3h$ excitations is
rather small on most of the computed excitation levels (except for the
second $1^+$ excited state that moves down by about $0.5$~MeV). For
$^{24}$Ne the agreement between EOM-CCSDT-1 and CCEI is overall
satisfactory. In particular the first $0^+$ state is in excellent
agreement, and the role of $3p$-$3h$ excitations is small. For the
first $2^+$ and $4^+$ states we see that $3p$-$3h$ are more important
and brings the EOM-CCSDT-1 result in closer agreement with CCEI. The
agreement with data for $^{24}$F and $^{24}$Ne is also quite good. For
$^{24}$F both CCEI and full-space coupled-cluster yield a
ground-state with spin and parity $J^\pi=3^+$ in agreement with
experiment \cite{caceres2015}.  Finally, we also compared our results
to those computed with in-medium SRG effective interactions and the
recent measurements of excited states in $^{24}$F \cite{caceres2015},
and found good agreement.

For the binding energies of $^{24}$F and $^{24}$Ne we obtain
$179.4$~MeV and $192.9$~MeV in CCEI, respectively, in good agreement
with the full-space coupled-cluster results of $181.0$~MeV and
$190.8$~MeV, respectively. Both CCEI and full-space coupled-cluster
results are in good agreement with the experimental binding energies
of $179.9$~MeV and $191.8$~MeV \cite{wang2012}. Finally we checked
that our calculations are reasonably well converged with respect to
the model-space size. In our full-space charge-exchange EOM-CCSDT-1
calculations we used $N_{\rm max}=12$ and an active space of $e_{3max}
= 12$ for the $3p$-$3h$ excitations for the ground-
and excited states in $^{24}$F, while for the excited states in
$^{24}$Ne we used $e_{3max} = 14$, and finally for the ground-state of
$^{24}$Ne we used $e_{3max} = 20$. We found that energies are
converged to within a few hundred keV with respect to these
active-space truncations. Beyond the active space truncation, there
are also uncertainties associated with the truncation of the
particle-hole excitation level in the EOM-CC approaches used to
compute the full-space charge exchange excitations (see
Fig.~\ref{FNe_benchmark}), and in the construction of the core-,
one-body, and two-body parts of the CCEI defined in
Eq.~\eqref{ham}. We refer the reader to \cite{jansen2014} for a more
detailed discussion on uncertainties related to the construction of
CCEI and the model-space truncations used.

\begin{figure}[tbh]
  \includegraphics[width=1.0\columnwidth]{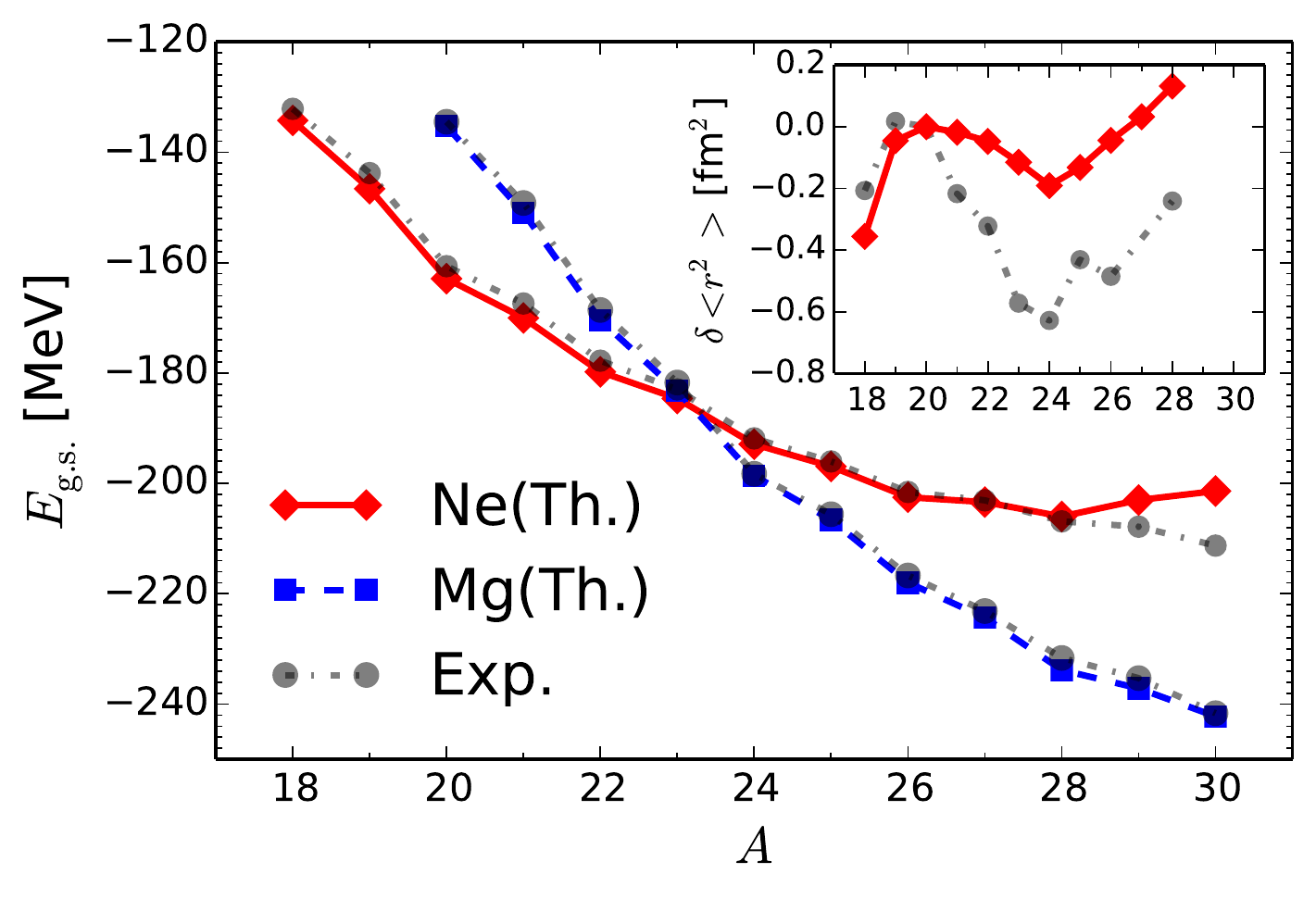}
  \caption{(Color online) Ground-state energies of neon (red line
    marked with diamonds) and magnesium isotopes (blue line marked
    with squares). Gray dashed-dotted lines marked with circles show
    the experimental values. The inset shows the CCEI results for the
    isotope shifts in neon isotopes, relative to $^{20}$Ne, compared to 
    known experimental data.}
  \label{Ne_Mg_BE_Rch}
\end{figure}

\begin{figure*}[tbh]
\includegraphics[width=1.0\textwidth]{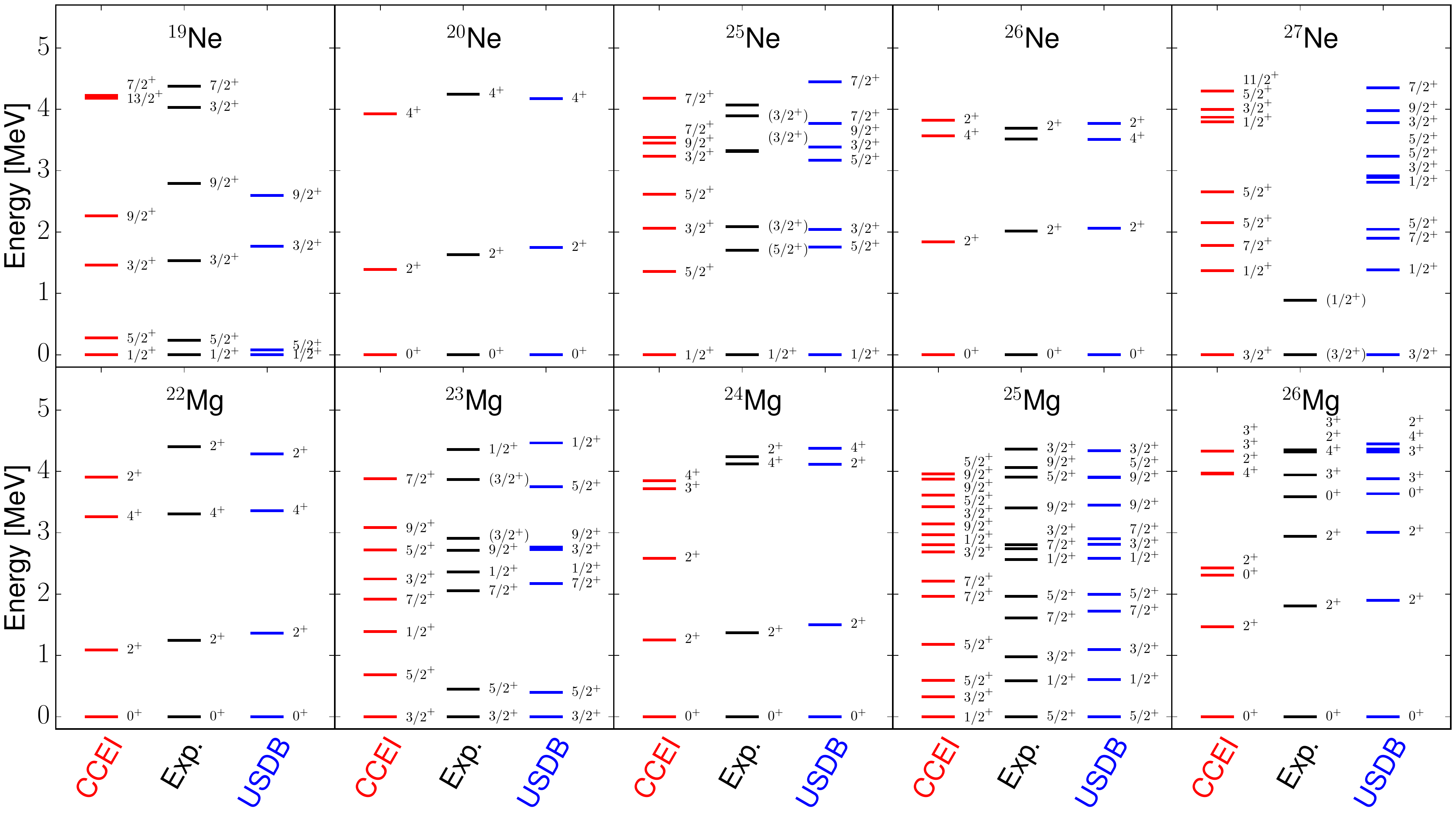}
\caption{(Color online) Excitation spectra of selected neon (upper
  panel) and magnesium (lower panel) isotopes. The left columns (red
  lines) display the CCEI results, the middle columns (black
  lines) include known positive-parity experimental states from the
  ENSDF database, while the right columns (blue lines) shows excitation spectra
  obtained from the USDB interaction~\cite{brown2006}.}
  \label{NeMg_panel}
\end{figure*}

Figure~\ref{Ne_Mg_BE_Rch} shows the total binding energies for
$^{18-30}$Ne and $^{20-30}$Mg obtained from our CCEI calculations and
are compared to data. We find a very good agreement between the CCEI
results and experiment for all magnesium isotopes and for neon
isotopes up to mass $A=28$. The deviation between the CCEI results and
experiment for $^{29,30}$Ne is not unexpected as these nuclei are part
of the well-known island of inversion region \cite{marinova2011}, for
which intruder states from the $fp$-shell become important. We also
computed binding energies for all isotope chains in the $sd$-shell and
found overall very good agreement with data (see Supplement
Material). In the inset, we show the computed isotope shifts of the
charge radii for the neon isotopes for which experimental data is
available \cite{angeli2013}. The CCEI calculations included the core
and one-body contributions to the radii, while the more demanding
inclusion of two-body contributions will be explored in the
future. The overall trend, in particular the kink at $N=24$, is
reproduced qualitatively.

In Fig. \ref{NeMg_panel}, we highlight the level schemes of a subset
of the computed neon and magnesium isotopes, including the
prototypical deformed nuclei $^{20}$Ne and $^{24}$Mg, as well as
odd-$A$ and neutron-rich exotic nuclei like $^{27}$Ne, for which
little experimental data is known.  We observe a quite reasonable
reproduction of the data in all cases, with a generally consistent
compression of the level scheme. We also observe rotational band
structures in these isotopic chains, as suggested from experimental
studies. 

The quality of these level schemes can be difficult to measure with
respect to experiment. To address the agreement with data and to
compare with the phenomenological USD-B interaction in a more
quantitative way we perform a comparison of the root-mean-squared
(rms) deviations obtained from 144 experimental levels in the lower
sd-shell. For the CCEI (USDB) shell-model interactions, we find values
of 591(244) keV for oxygen, 452(268) keV in fluorine, 422(268) keV in
neon, 529(155) keV in sodium and 760(106) keV magnesium. We note that
the CCEI rms deviations are very close to the corresponding IM-SRG rms
deviations reported in \textcite{stroberg16}.

\begin{figure}[tbh]
  \includegraphics[width=1.0\columnwidth]{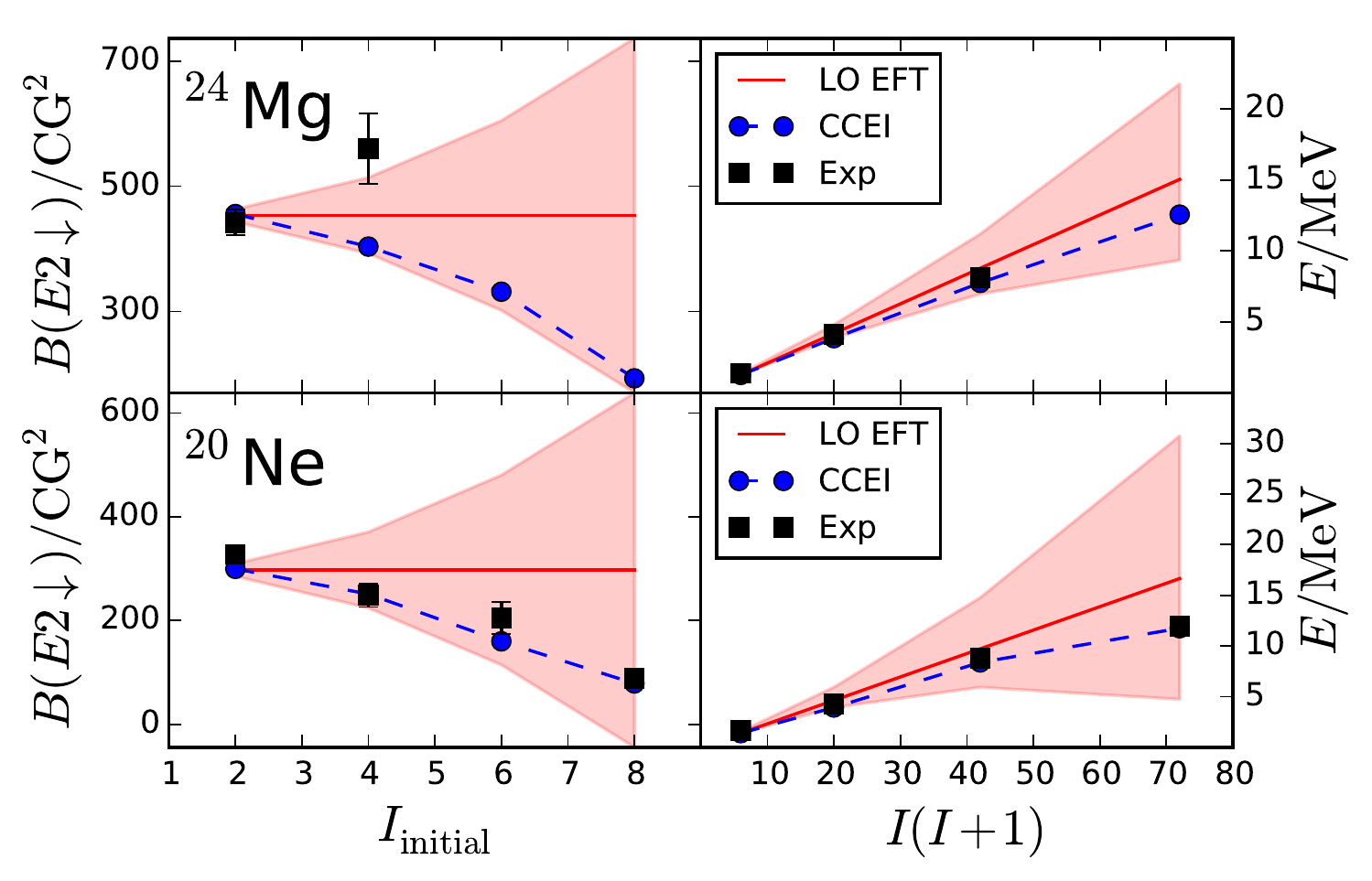}
  \caption{(Color online) $B(E2\downarrow)$ transition strengths,
    divided by a squared Clebsch Gordan coefficient CG, as a function
    of initial spin (left panels) and corresponding excitation
    energies as a function of spin (right panels) for $^{24}$Mg (top
    panels) and $^{20}$Ne (bottom panels). Data (black squares) and
    results from CCEI (blue circles connected by dashed lines) are
    shown together with results from a leading-order EFT for deformed
    nuclei (red line and shaded uncertainty estimates).}
  \label{panel}
\end{figure}

Let us now turn to the deformed nuclei $^{20}$Ne and $^{24}$Mg in more
detail.  Deformed even-even nuclei exhibit ground-state rotational
bands with energies $E(I)=\alpha_0 I(I+1)[1+\alpha_2 I(I+1)
+\ldots]$~\cite{bohr1952,bohr1953,mariscotti1969}.  Here,
$I=0,2,4,\ldots$ denotes the spin, $\alpha_0$ is the rotational
constant (i.e. twice the inverse of the moment of inertia), and
$\alpha_2\ll 1$ is a small correction. Similarly, strengths of
electric quadrupole transitions $B(E2\downarrow,I_i)=Q_0^2 ({\rm
  CG})^2 [1+\beta_2 I_i(I_i-1)+\ldots]$ from an initial spin $I_i$ to
final spin $I_i-2$ can be expanded as a function of
spin~\cite{mikhailov1964,mikhailov1966}. Here ${\rm CG}$ denotes a
Clebsch Gordan coefficient that accounts for geometric aspects in the
transition, while $Q_0$ and $\beta_2\ll 1$ are low-energy
constants. For a rigid rotor, $\alpha_2=0=\beta_2$. The relations for
energies and $B(E2)$ transitions, well known from collective models of
the atomic nucleus~\cite{eisenberg1970,bohr1975,iachello1987}, were
recently re-derived via
EFT~\cite{papenbrock2011,papenbrock2014,coelloperez2015}.  The EFT
exploits the separation of scale between the energy $\xi$ of rotations
and the breakdown energy $\Omega$ that marks the proliferation of
non-rotational degrees of freedom. For even-even $sd$-shell nuclei
$\xi\approx 1.5$~MeV, and $\Omega\approx 6$~MeV. The low-energy
constants $\alpha_2$ and $\beta_2$ that account for deviations from
the rigid rotor are of the order $(\xi/\Omega)^2$.  Figure~\ref{panel}
shows that computed $B(E2)$ transitions and spectra agree well with
data for $^{20}$Ne and $^{24}$Mg. Note that for $B(E2)$ we used the
standard effective charges of $e_{\rm eff}^{\rm p}=1.5e$ and $e_{\rm
  eff}^{\rm n}=0.5e$, for protons and neutrons, respectively. We
verified that the results shown for the $B(E2)$ transitions are not
sensitive to small variations of the effective charge. Furthermore, in
\textcite{navratil1997} it was shown that these phenomenological 
effective charges are close to fully microscopically derived effective 
charges for the $p$-shell. In agreement with EFT results, the 
deviations from the rigid-rotor behavior (red solid lines) are of similar 
size for spectra and transitions, and within EFT uncertainty estimates 
(red shaded regions).

{\it Conclusions} -- In this work we extended the \emph{ab initio}
CCEI method to deformed $sd$-shell nuclei. We presented results for
binding energies and excited states in isotopes of neon and magnesium
based on chiral nucleon-nucleon and three-nucleon forces. The results
were found to be in good agreement with data. Our 
calculations reproduce rotational bands and $B(E2)$ transitions in
$^{20}$Ne and $^{24}$Mg within the uncertainties estimated from an EFT
derived for deformed nuclei. We have thus extended the description of
collective degrees of freedom as emergent phenomena in $sd$-shell
nuclei from first principles. This work paves the way for
non-perturbative shell-model Hamiltonians tied to chiral EFT, with
predictive power in the $sd$-shell and beyond.

\begin{acknowledgments}
  We would in particular like to thank Ragnar Stroberg for computing
  the rms deviations from experiment for our CCEI results and for
  providing us with shell-model results for the $sd$-shell nuclei
  considered in this work using the USD-B interaction. We also thank
  Heiko Hergert, Jason Holt, Jonathan Engel and Thomas Papenbrock for
  useful discussions. We are particularly grateful to Thomas
  Papenbrock for generating Figure 4 and contributing to the
  discussion related to the effective field theory for deformed
  nuclei. This work was supported by the Office of Nuclear Physics,
  U.S. Department of Energy (Oak Ridge National Laboratory),
  DE-SC0008499 (NUCLEI SciDAC collaboration), NERRSC Grant No.\
  491045-2011, and the Field Work Proposal ERKBP57 at Oak Ridge
  National Laboratory.  Computer time was provided by the Innovative
  and Novel Computational Impact on Theory and Experiment (INCITE)
  program.  TRIUMF receives funding via a contribution through the
  National Research Council Canada. This research used resources of
  the Oak Ridge Leadership Computing Facility located in the Oak Ridge
  National Laboratory, which is supported by the Office of Science of
  the Department of Energy under Contract No.  DE-AC05-00OR22725, and
  used computational resources of the National Center for
  Computational Sciences and the National Institute for Computational
  Sciences.
\end{acknowledgments}

\bibliography{prl_resub}
\bibliographystyle{apsrev}
    
\end{document}

\begin{figure}[t]
  \includegraphics[width=1.0\columnwidth]{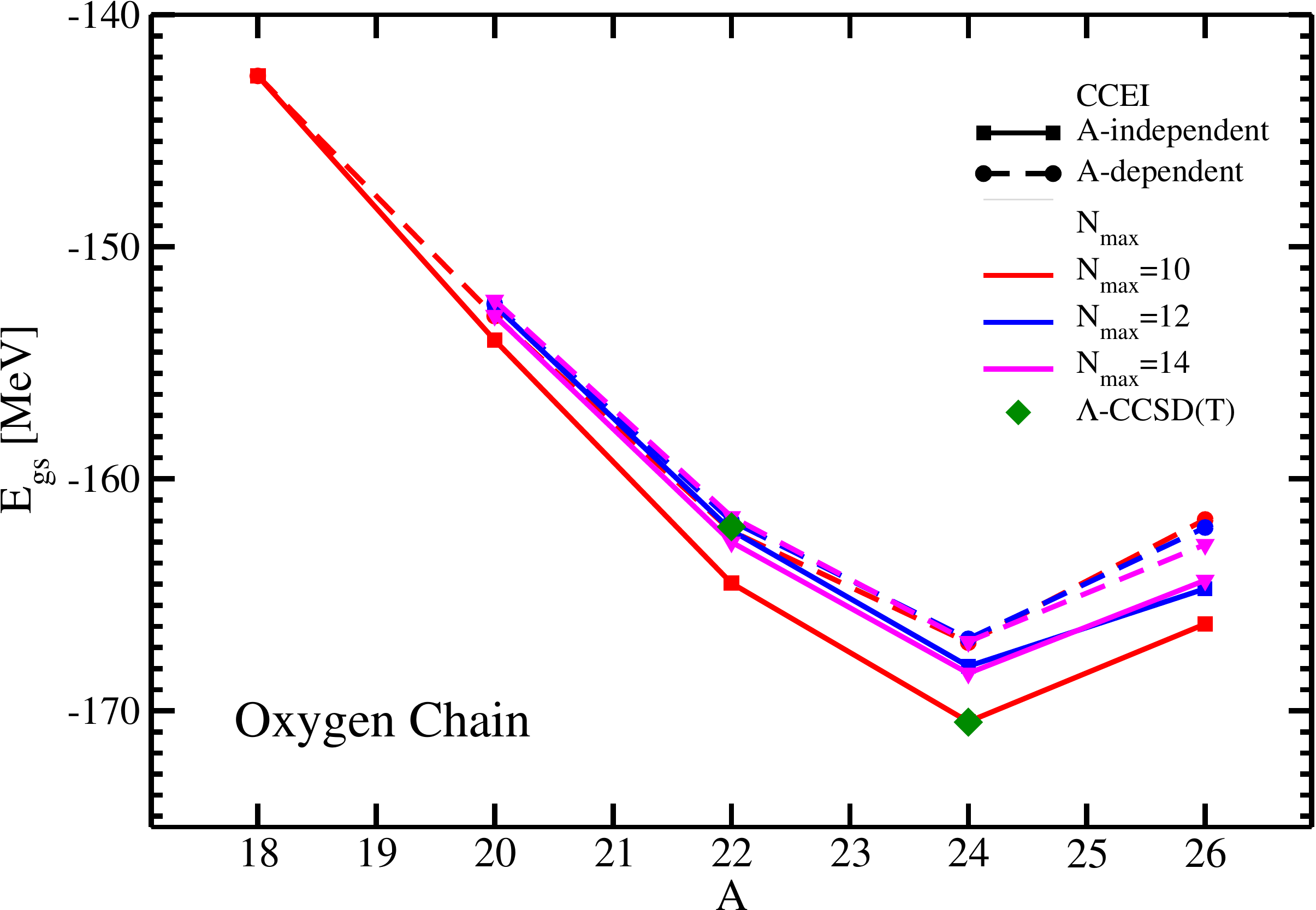}
  \caption{(Color online) Ground state energies of the oxygen chain
    using two constructions of the CCEI: $A$-independent (solid lines)
    and $A$-dependent (dashed lines) (see text for description),
    computed using a coupled cluster model space of $N_\text{max}=10$
    (red), $12$ (blue) and $14$ (magenta) compared to full space
    $\Lambda$-CCSD(T) (diamonds). The calculations were performed
    using the chiral nucleon-nucleon and three-nucleon interaction
    used in this work.}
  \label{Adependence}
\end{figure}

{\it Supplementary material.} -- When constructing the CCEI one can
choose the mass $A$ that appears in the intrinsic kinetic energy of
the individual parts of the valence cluster expansion in two ways: (i)
by the mass of the target nucleus or (ii) by the mass of the core,
one-body and two-body parts. This freedom has a relatively large
impact on the computation time and, as we will see below, less so on
the binding energy and excited states of the target
nucleus. Computationally, the former method requires the calculation
of the CCEI for each desired $A$, whereas the latter choice only needs
one CCEI calculation which can then be used for any nucleus in the
valence space. In Fig.~\ref{Adependence} we show the computed CCEI
ground-state energies for selected oxygen isotopes using a model-space
size of $N_\text{max}=10$, $12$, and $14$ and the chiral
nucleon-nucleon and three-nucleon force used in this work. The CCEI
results are then compared to the full-space $\Lambda$-CCSD(T) results.

\begin{figure}[t]
  \includegraphics[width=1.0\columnwidth]{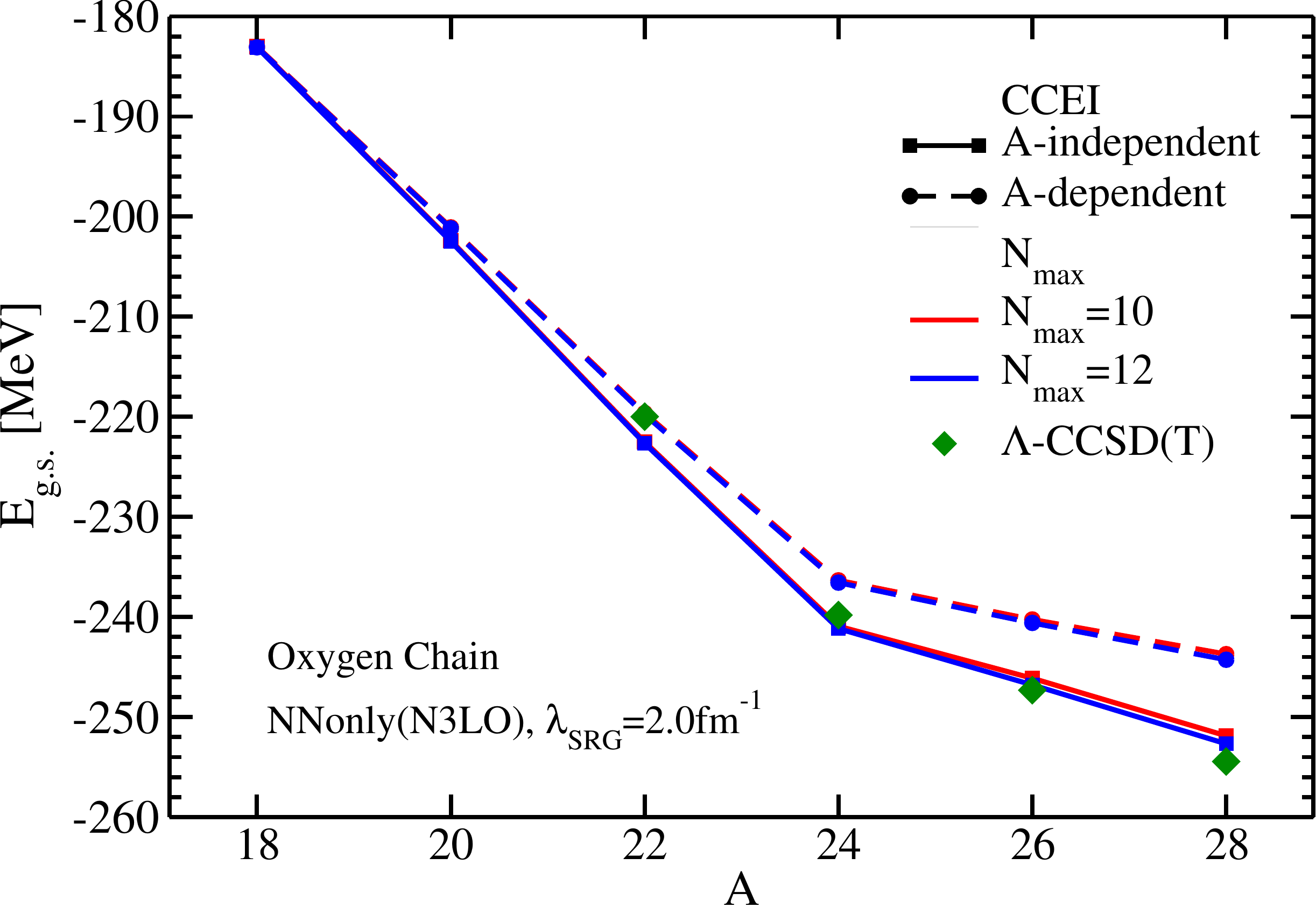}
  \caption{(Color online) Ground state energies of the oxygen chain
    using two constructions of the CCEI: $A$-independent (solid lines)
    and $A$-dependent (dashed lines) (see text for description),
    computed using a coupled cluster model space of $N_\text{max}=10$
    (red) and $12$ (blue) compared to full space $\Lambda$-CCSD(T)
    (diamonds). The calculations were performed using the chiral
    nucleon-nucleon interaction N$^3$LO of \textcite{entem2003}
    evolved via SRG to $\lambda=2.0\text{fm}^{-1}$.}
  \label{Adependence_NNonly}
\end{figure}

The ground state energy calculations are converged to less than
$0.4\%$ with respect to the model-space used to generate the CCEI at
$N_\text{max}=14$. The ground-state energies computed using the
$A$-dependent and $A$-independent CCEI approach converge as the size
of the model space is increased. We observe that the $A$-independent
CCEI results track closer to the full-space $\Lambda$-CCSD(T) results
than the corresponding $A$-dependent CCEI results. We note that for
the results presented in Fig.~\ref{Adependence} the full-space
coupled-cluster method uses a Hamiltonian that is normal-ordered with
respect to the reference states of $^{22,24,28}$O, respectively. For
CCEI method we normal-order the Hamiltonian with respect to
$^{16}$O. As we are using the normal-ordered two-body approximation
and thereby neglect the residual normal-ordered three-body parts of
the Hamitlonian, one might argue that the full-space coupled-cluster
and CCEI results for $^{22,24,28}$O are based on slightly different
Hamiltonians. In order to remove the effects of the normal ordering
with respect to different reference states in our $A$-dependence
investigation of the CCEI, we also computed the oxygen chain using
only the nucleon-nucleon interaction of our Hamiltonian, shown in
Figure \ref{Adependence_NNonly}.

\begin{figure}[t]
  \includegraphics[width=1.0\columnwidth]{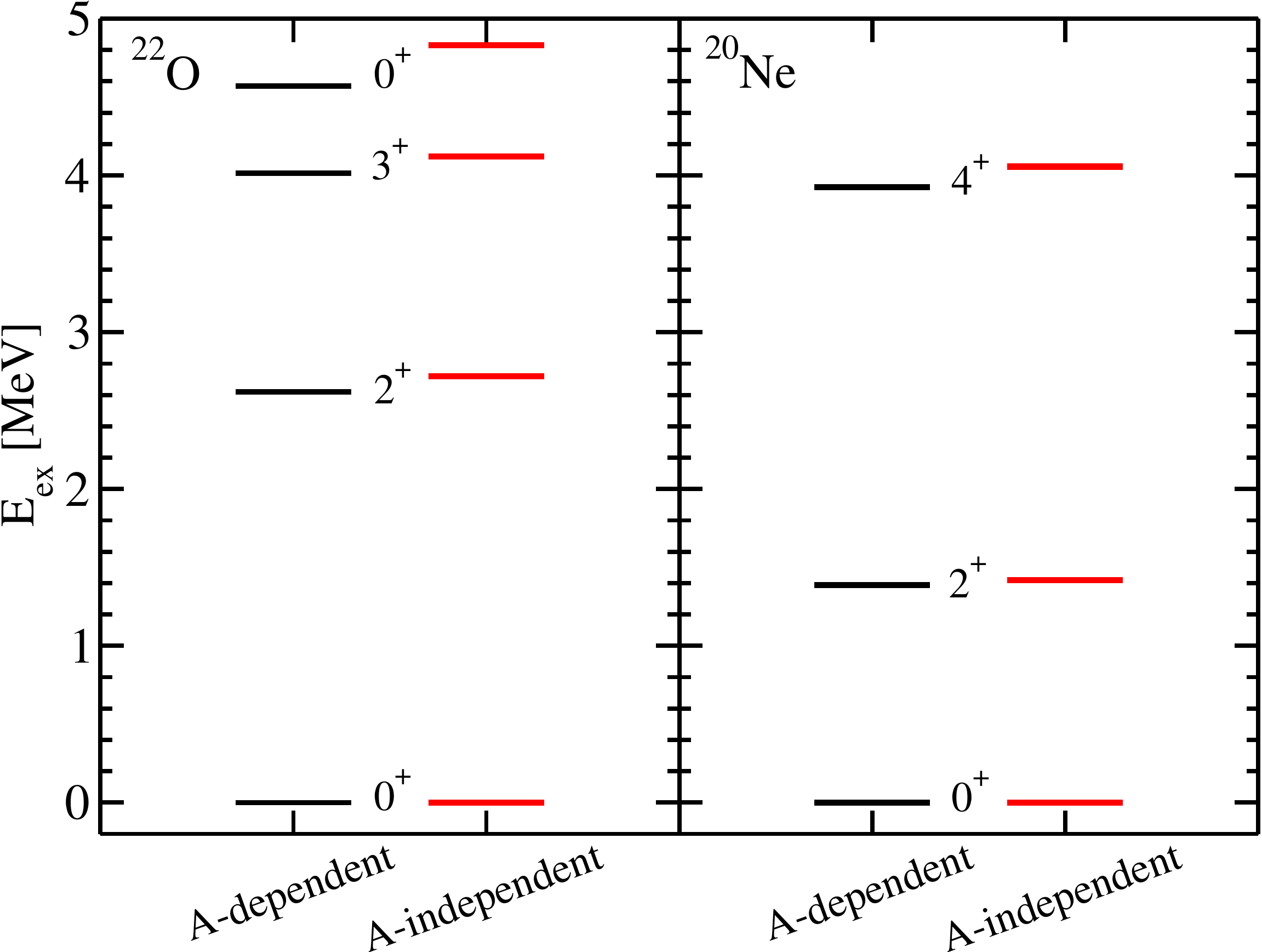}
  \caption{(Color online) Spectra of $^{22}$O (left panel) and
    $^{20}$Ne (right panel) using two constructions of the CCEI:
    $A$-independent and $A$-dependent (see text for description).  The
    calculations were performed using the chiral nucleon-nucleon and
    three-nucleon interaction used in this work.}
  \label{Adependence_spectra}
\end{figure}

Here we generate the CCEI using the $A$-independent and $A$-dependent
constructions at $N_\text{max}$=10 and 12 and compare them to full
space $\Lambda$-CCSD(T) calculations. The $A$-independent construction
tracks the full space calculations even better than when we include
the three-body force. This helps to strengthen our choice in using the
$A$-independent CCEI as our preferred
method. Figure~\ref{Adependence_spectra} shows the spectra of $^{22}$O
and $^{20}$Ne for both the $A$-independent and $A$-dependent
interactions.

\begin{figure*}[h!]
  \includegraphics[width=1.0\columnwidth]{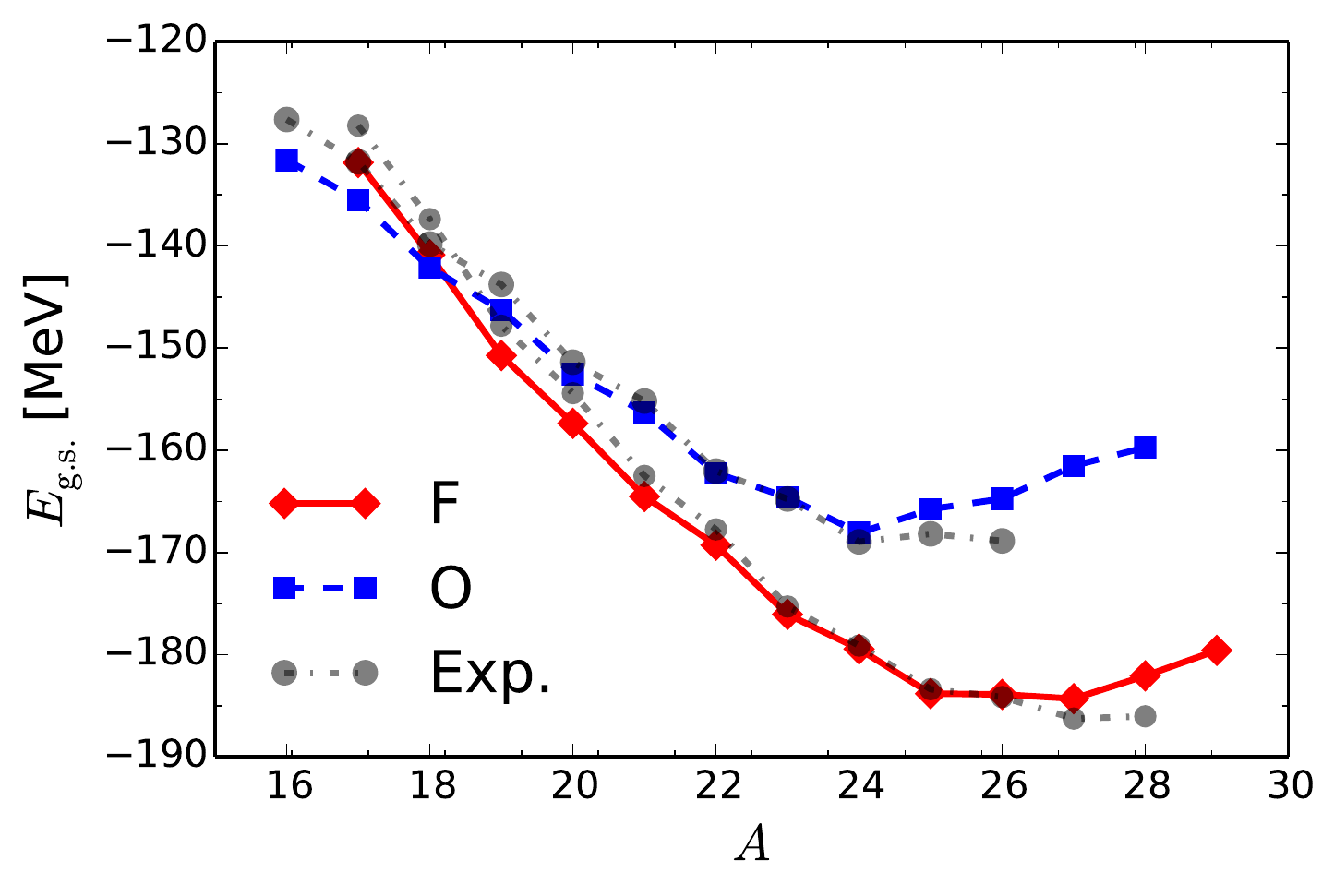}
  \includegraphics[width=1.0\columnwidth]{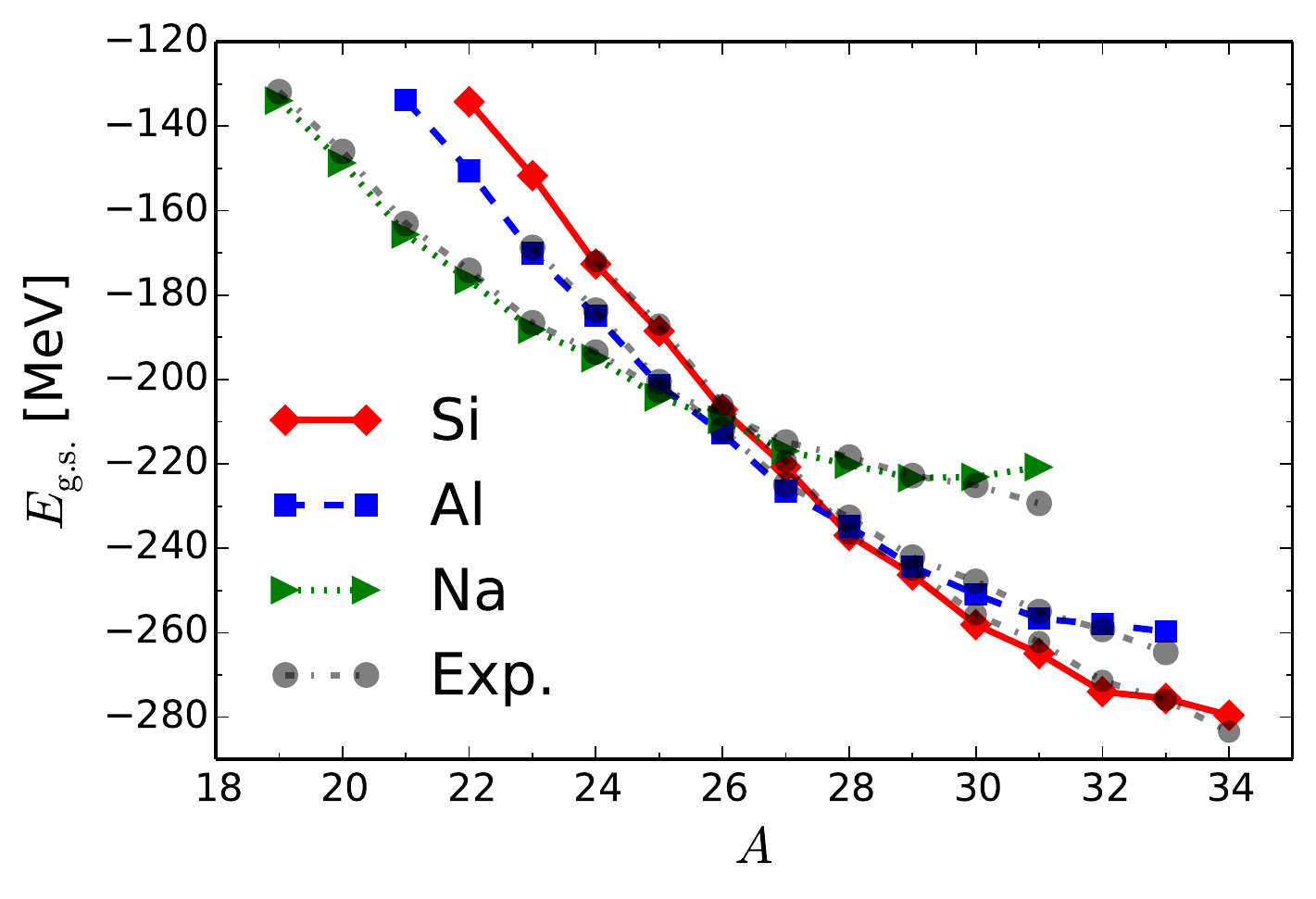}
   \includegraphics[width=1.0\columnwidth]{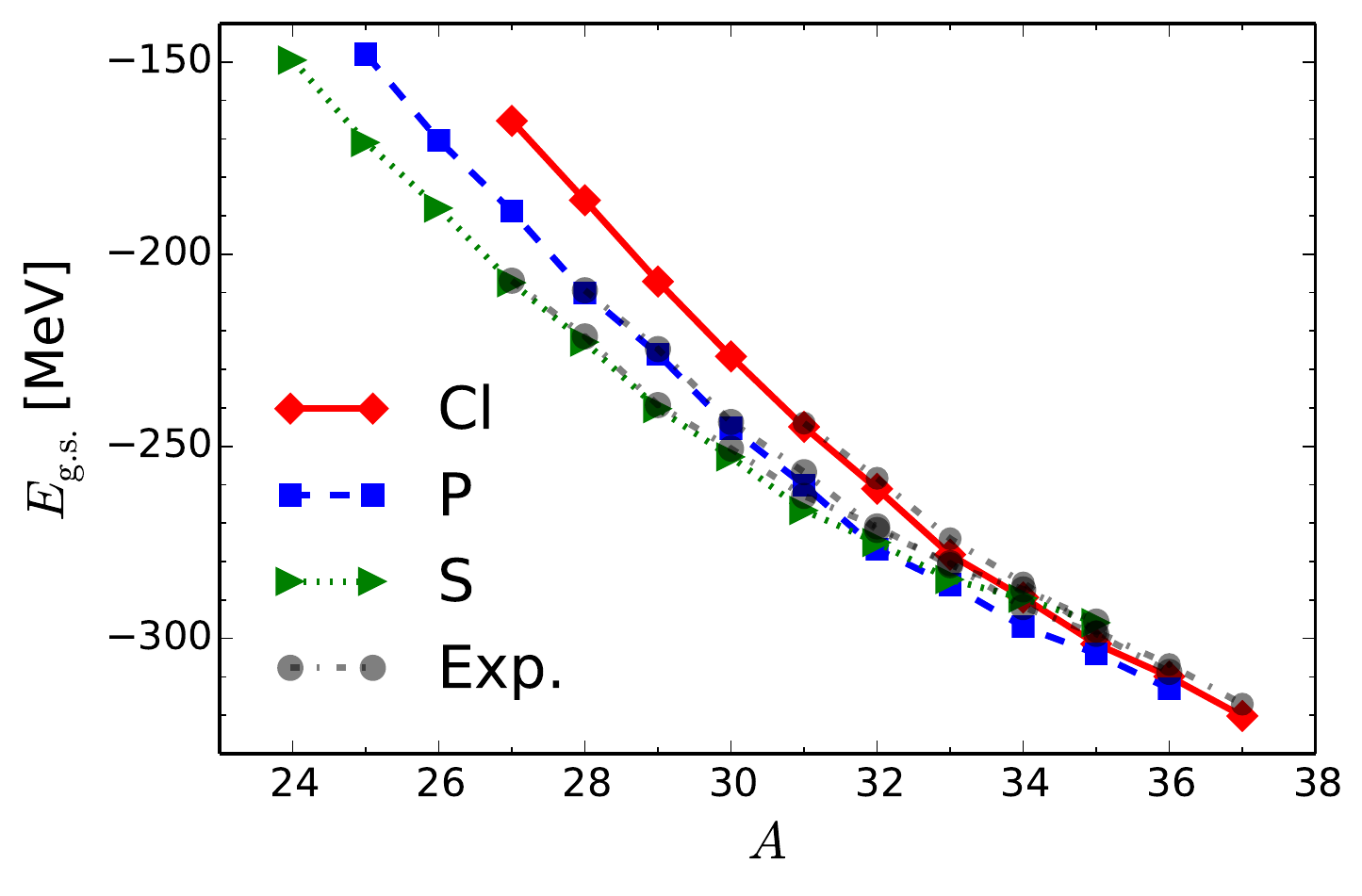}  
  \includegraphics[width=1.0\columnwidth]{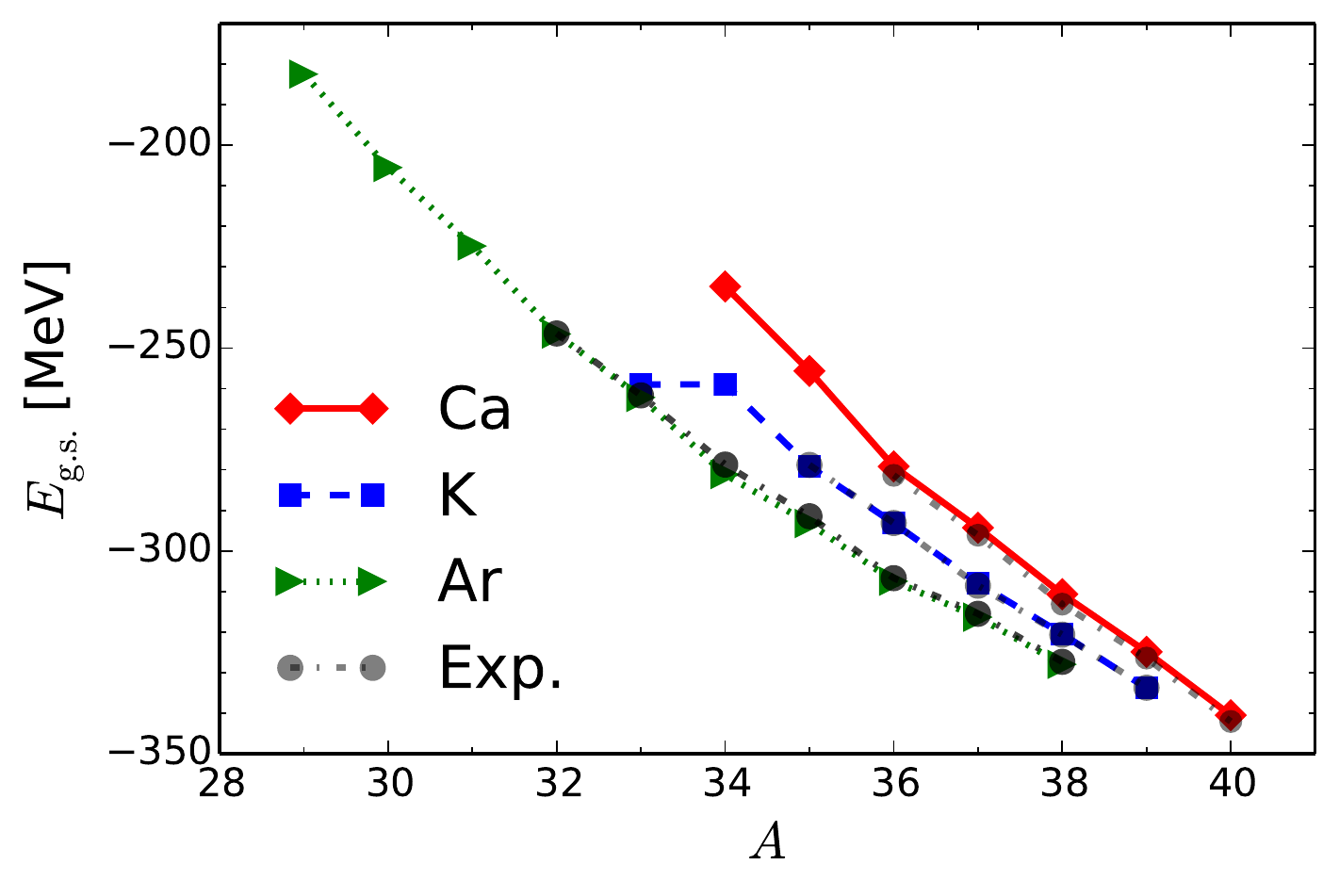}
  \caption{(Color online) Upper left: Ground-state energies of
    fluorine (solid line marked with diamonds) and oxygen isotopes
    (dash-dotted line marked with squares). Upper right: Ground-state
    energies of silisium (solid line marked with diamonds), aluminium
    (dash-dotted line marked with squares), and natrium isotopes
    (dotted line with right triangles). Lower left: Ground-state
    energies of chlorine (solid line marked with diamonds), phosphorus
    (dash-dotted line marked with squares), and sulfur isotopes
    (dotted line with right triangles) Lower right: Ground-state
    energies of calcium (solid line marked with diamonds), potassium
    (dash-dotted line marked with squares), and argon isotopes (dotted
    line with right triangles). Dashed-dotted lines marked with
    circles show the experimental values. Dashed-dotted lines marked
    with circles show the experimental values.}
  \label{lower_sd}
\end{figure*}

The change in spectra between the two CCEI interactions is generally
small, with an average shift of approximately 100 keV. This is
comparable to the average shift due to the change in model-space
size. The largest shifts tend appear in the higher lying states,
e.g. the approximately 200 keV shift in the second $0^+$ state in
$^{22}$O and the 150 keV shift in the $4^+$ state in $^{20}$Ne. Since
the difference between the two constructions of the CCEI has the
largest affect on the binding energies, we choose the $A$-independent
interaction as our preferred method. Not only is it easier from a
computational perspective, it also tends to give the best results when
compared to full-space coupled-cluster calculations considered here.
Figure~\ref{lower_sd} shows the binding energies for isotope chains in
the \emph{sd}-shell obtained in our CCEI approach and compared to
experimental data.

The CCEI shell-model Hamiltonian in the two-body valence-cluster
approximation is given by 
\begin{equation}
  \label{ham}
  H_{\rm CCEI} = H_0^{A_c}+ H_1^{A_c+1} + H_2^{A_c+2},
\end{equation}
here $H_0^{A_c}$ is the the core ground-state energy, $H_1^{A_c+1} $
is the one-body part and $H_2^{A_c+2}$ the two-body part. In this work
the $^{16}$O core energy computed in the $\Lambda$-CCSD(T)
approximation is $-131.39$~MeV. The diagonal one-body matrix elements
$\langle a \vert H_1^{A_c+1}\vert a\rangle $ for the $sd$-shell is
given in Table~\ref{tab3}, while the two-body matrix matrix elements
$\langle ab\vert H_2^{A_c+2} \vert cd\rangle $ are given in
Tables~\ref{tab1} and~\ref{tab2}.

\begin{table}[b]
 \begin{ruledtabular}
     \begin{tabular}{|ccccc|c|}
  $a$ & $n_a$ & $l_a$ &  $2j_a $ & $2t_{z_a}$ &  $\langle a \vert H_1^{A_c+1}\vert a\rangle $ \\ \hline
 1  & 0    &  2   &     3   & -1   &     7.05777400 \\
 2  & 0   &   2  &      5  &  -1  &   -0.25337698 \\
 3  & 1    &  0   &     1   & -1   &     1.10565334 \\ 
 4  & 0    &  2   &     3   &  1    &     3.85759640 \\
 5  & 0    &  2   &     5   &  1    &   -3.93783920 \\
 6  & 1    &  0   &     1   &  1    &   -2.25063950 \\
    \end{tabular}
 \end{ruledtabular}
   \caption{Single-particle basis and one-body matrix-elements for the $sd$-shell from
     CCEI. $a$ labels the single-particle orbit with corresponding
     quantum numbers $n_al_aj_at_{z_a}$ (nodal number, orbital angular
     momentum, total angular momentum, and isospin projection $t_Z =
     -1/2$ for protons and $t_z =1/2$ for neutrons ).}
   \label{tab3}
 \end{table}

\begin{table*}[b]
\begin{ruledtabular}
    \begin{tabular}{|llllll|r|llllll|r|}
     $a $ &$b $ &$c $ &$d $ & $J$ & $T$ & $\langle ab\vert
     H_2^{A_c+2}\vert cd\rangle$ & $a $ &$b $ &$c $ &$d $ & $J$ & $T$ & $\langle ab\vert H_2^{A_c+2}\vert cd\rangle$ \\\hline
  2  &  2  &  2  &  2  &   0  &  1  &  -1.209884          &            1   & 5  &  1  &  5  &   1  &  0  &  -4.732176   \\        
  2  &  2  &  3  &  3  &   0  &  1  &  -1.267748          &            1   & 5  &  3  &  6  &   1  &  0  &  -2.634041   \\        
  2  &  2  &  1  &  1  &   0  &  1  &  -2.813678          &            1   & 5  &  3  &  4  &   1  &  0  &   1.417620    \\       
  3  &  3  &  3  &  3  &   0  &  1  &  -0.789697          &            1   & 5  &  1  &  6  &   1  &  0  &  -1.600826   \\          
  3  &  3  &  1  &  1  &   0  &  1  &  -0.627860          &            1   & 5  &  1  &  4  &   1  &  0  &  -0.401344   \\        
  1  &  1  &  1  &  1  &   0  &  1  &  -0.463086          &            3   & 6  &  3  &  6  &   1  &  0  &  -2.735304   \\          
  2  &  1  &  2  &  1  &   1  &  1  &   0.508993          &            3   & 6  &  3  &  4  &   1  &  0  &   0.063961     \\       
  2  &  1  &  3  &  1  &   1  &  1  &  -0.145290          &            3   & 6  &  1  &  6  &   1  &  0  &  -0.004960   \\        
  3  &  1  &  3  &  1  &   1  &  1  &   0.471263          &            3   & 6  &  1  &  4  &   1  &  0  &  -0.326177   \\        
  2  &  2  &  2  &  2  &   2  &  1  &  -0.354285          &            3   & 4  &  3  &  4  &   1  &  0  &  -4.431654   \\        
  2  &  2  &  2  &  3  &   2  &  1  &  -0.780041          &            3   & 4  &  1  &  6  &   1  &  0  &   4.876612      \\        
  2  &  2  &  2  &  1  &   2  &  1  &  -0.354356          &            3   & 4   & 1  &  4  &   1  &  0  &   2.230581      \\        
  2  &  2  &  3  &  1  &   2  &  1  &  -0.752583          &            1   & 6   & 1  &  6  &   1  &  0  &  -4.488804       \\       
  2  &  2  &  1  &  1  &   2  &  1  &  -0.604916          &            1   & 6   & 1  &  4  &   1  &  0  &  -2.255396      \\        
  2  &  3  &  2  &  3  &   2  &  1  &  -0.540280          &            1   & 4   & 1  &  4  &   1  &  0  &  -1.443965      \\        
  2  &  3  &  2  &  1  &   2  &  1  &  -0.200114          &            2   & 5   & 2  &  5  &   2  &  0  &  -0.748201      \\        
  2  &  3  &  3  &  1  &   2  &  1  &  -1.191250          &            2   & 5   & 2  &  6  &   2  &  0 &  -0.726085       \\       
  2  &  3  &  1  &  1  &   2  &  1  &  -0.394891          &            2   & 5   & 2  &  4  &   2  &  0  &  -0.272280      \\        
  2  &  1  &  2  &  1  &   2  &  1  &   0.503484          &            2   & 5   & 3  &  5  &   2  &  0  &  -0.716467      \\        
  2  &  1  &  3  &  1  &   2  &  1  &  -0.480941          &            2   & 5   & 1  &  5  &   2  &  0  &   0.398931      \\        
  2  &  1  &  1  &  1  &   2  &  1  &  -0.747132          &            2   & 5   & 3  &  4  &   2  &  0  &  -0.774583     \\         
  3  &  1  &  3  &  1  &   2  &  1  &  -0.144091          &            2   & 5   & 1  &  6  &   2  &  0  &   0.667225     \\         
  3  &  1  &  1  &  1  &   2  &  1  &  -0.240292          &            2   & 5   & 1  &  4  &   2  &  0  &  -1.247244      \\        
  1  &  1  &  1  &  1  &   2  &  1  &   0.317194          &            2   & 6   & 2  &  6  &   2  &  0  &  -1.676750     \\         
  2  &  3  &  2  &  3  &   3  &  1  &   0.835428          &            2   & 6   & 2  &  4  &   2  &  0  &  -2.194518     \\         
  2  &  3  &  2  &  1  &   3  &  1  &  -0.046666          &            2   & 6   & 3  &  5  &   2  &  0  &   0.132552     \\         
  2  &  1  &  2  &  1  &   3  &  1  &   0.539996          &            2   & 6   & 1  &  5  &   2  &  0  &  -2.101734      \\        
  2  &  2  &  2  &  2  &   4  &  1  &   0.384441          &            2   & 6   & 3  &  4  &   2  &  0  &  -3.530858     \\         
  2  &  2  &  2  &  1  &   4  &  1  &  -1.166757          &            2   & 6   & 1  &  6  &   2  &  0  &  -0.573096     \\         
  2  &  1  &  2  &  1  &   4  &  1  &  -0.618327          &            2   & 6   & 1  &  4  &   2  &  0  &  -1.753591     \\         
  5   & 5  &  5  &  5  &   0  &  1  &  -1.683538          &            2   & 4   & 2  &  4  &   2  &  0  &  -4.072754      \\        
  5   & 5  &  6  &  6  &   0  &  1  &  -1.306942          &            2   & 4   & 3  &  5  &   2  &  0  &   2.186536      \\        
  5   & 5  &  4  &  4  &   0  &  1  &  -3.025528          &            2   & 4   & 1  &  5  &   2  &  0  &  -4.432840       \\       
  6   & 6  &  6  &  6  &   0  &  1  &  -1.176083          &            2   & 4   & 3  &  4  &   2  &  0  &  -2.600212      \\        
  6   & 6  &  4  &  4  &   0  &  1  &  -1.100067          &            2   & 4   & 1  &  6  &   2  &  0  &  -1.389508      \\        
  4   & 4  &  4  &  4  &   0  &  1  &  -0.774292          &            2   & 4   & 1  &  4  &   2  &  0  &  -1.303326      \\        
  5   & 4  &  5  &  4  &   1  &  1  &   0.219121          &            3   & 5   & 3  &  5  &   2  &  0  &  -1.626130       \\       
  5   & 4  &  6  &  4  &   1  &  1  &  -0.149186          &            3   & 5   & 1  &  5  &   2  &  0  &   2.507310      \\        
  6   & 4  &  6  &  4  &   1  &  1  &   0.225270          &            3   & 5   & 3  &  4  &   2  &  0  &   0.903296      \\        
    \end{tabular}                                                                                                                          
  \end{ruledtabular}
  \caption{CCEI two-body matrix-elements for the $sd$-shell.}
  \label{tab1}
\end{table*}            

\begin{table*}[b]
\begin{ruledtabular}
    \begin{tabular}{|llllll|r|llllll|r|}
     $a $ &$b $ &$c $ &$d $ & $J$ & $T$ & $\langle ab\vert
     H_2^{A_c+2}\vert cd\rangle$ & $a $ &$b $ &$c $ &$d $ & $J$ & $T$ & $\langle ab\vert H_2^{A_c+2}\vert cd\rangle$ \\\hline
 5   & 5  &  5  &  5  &   2  &  1  &  -0.788439          &            3   & 5   & 1  &  6  &   2  &  0  &   2.583444      \\        
  5   & 5  &  5  &  6  &   2  &  1  &  -0.848394          &            3   & 5   & 1  &  4  &   2  &  0  &  -2.040815      \\        
  5   & 5  &  5  &  4  &   2  &  1  &  -0.375108          &            1   & 5   & 1  &  5  &   2  &  0  &  -4.207750      \\        
  5   & 5  &  6  &  4  &   2  &  1  &  -0.837176          &            1   & 5   & 3  &  4  &   2  &  0  &  -1.365998       \\       
  5   & 5  &  4  &  4  &   2  &  1  &  -0.580298          &            1   & 5   & 1  &  6  &   2  &  0  &  -2.148594      \\        
  5   & 6  &  5  &  6  &   2  &  1  &  -0.949096          &            1   & 5   & 1  &  4  &   2  &  0  &   0.007608      \\        
  5   & 6  &  5  &  4  &   2  &  1  &  -0.218504          &            3   & 4   & 3  &  4  &   2  &  0  &  -2.661450      \\        
  5   & 6  &  6  &  4  &   2  &  1  &  -1.341486          &            3   & 4   & 1  &  6  &   2  &  0  &  -2.097366       \\       
  5   & 6  &  4  &  4  &   2  &  1  &  -0.372504          &            3   & 4   & 1  &  4  &   2  &  0  &  -0.420498      \\        
  5   & 4  &  5  &  4  &   2  &  1  &   0.191078          &            1   & 6   & 1  &  6  &   2  &  0  &  -3.186614      \\        
  5   & 4  &  6  &  4  &   2  &  1  &  -0.552507          &            1   & 6   & 1  &  4  &   2  &  0  &   1.263075      \\        
  5   & 4  &  4  &  4  &   2  &  1  &  -0.782881          &            1   & 4   & 1  &  4  &   2  &  0  &  -0.633272       \\       
  6   & 4  &  6  &  4  &   2  &  1  &  -0.483819          &            2   & 5   & 2  &  5  &   3  &  0  &  -1.254326       \\       
  6   & 4  &  4  &  4  &   2  &  1  &  -0.205074          &            2   & 5   & 2  &  6  &   3  &  0  &  -1.985546        \\      
  4   & 4  &  4  &  4  &   2  &  1  &   0.091233          &            2   & 5   & 2  &  4  &   3  &  0  &   2.550492       \\       
  5   & 6  &  5  &  6  &   3  &  1  &   0.557494          &            2   & 5   & 3  &  5  &   3  &  0  &  -2.065423       \\       
  5   & 6  &  5  &  4  &   3  &  1  &  -0.061753          &            2   & 5   & 1  &  5  &   3  &  0  &  -2.607278       \\       
  5   & 4  &  5  &  4  &   3  &  1  &   0.266798          &            2   & 5   & 1  &  4  &   3  &  0  &  -0.091707        \\      
  5   & 5  &  5  &  5  &   4  &  1  &   0.017251          &            2   & 6   & 2  &  6  &   3  &  0  &  -3.719068       \\       
  5   & 5  &  5  &  4  &   4  &  1  &  -1.289670          &            2   & 6   & 2  &  4  &   3  &  0  &   1.264112       \\       
  5   & 4  &  5  &  4  &   4  &  1  &  -1.041445          &            2   & 6   & 3  &  5  &   3  &  0  &  -4.705594       \\       
  2   & 5  &  2  &  5  &   0  &  0  &  -1.835201          &            2   & 6   & 1  &  5  &   3  &  0  &  -1.304042      \\        
  2   & 5  &  3  &  6  &   0  &  0  &  -1.311656          &            2   & 6   & 1  &  4  &   3  &  0  &  -0.163411      \\        
  2   & 5  &  1  &  4  &   0  &  0  &  -3.023355          &            2   & 4   & 2  &  4  &   3  &  0  &   2.816222       \\       
  3   & 6  &  3  &  6  &   0  &  0  &  -1.209983          &            2   & 4   & 3  &  5  &   3  &  0  &   0.595824      \\        
  3   & 6  &  1  &  4  &   0  &  0  &  -1.087112          &            2   & 4   & 1  &  5  &   3  &  0  &  -3.405766      \\        
  1   & 4  &  1  &  4  &   0  &  0  &  -0.929649          &            2   & 4   & 1  &  4  &   3  &  0  &  -0.304057      \\        
  2   & 5  &  2  &  5  &   1  &  0  &  -1.895294          &            3   & 5   & 3  &  5  &   3  &  0  &  -3.428880      \\       
  2   & 5  &  2  &  4  &   1  &  0  &   3.680259          &            3   & 5   & 1  &  5  &   3  &  0  &  -0.117532      \\        
  2   & 5  &  1  &  5  &   1  &  0  &  -3.479575          &            3   & 5   & 1  &  4  &   3  &  0  &   0.732445      \\        
  2   & 5  &  3  &  6  &   1  &  0  &  -1.103603          &            1   & 5   & 1  &  5  &   3  &  0  &   5.259282      \\        
  2   & 5  &  3  &  4  &   1  &  0  &  -0.124722          &            1   & 5   & 1  &  4  &   3  &  0  &   0.951971       \\       
  2   & 5  &  1  &  6  &   1  &  0  &   0.148509          &            1  &  4   & 1   & 4   &  3   & 0 & -1.947045         \\       
  2   & 5  &  1  &  4  &   1  &  0  &   1.543712          &            2  &  5   & 2   & 5   &  4   & 0 & -0.014895          \\      
  2   & 4  &  2  &  4  &   1  &  0  &  -5.374974          &            2  &  5   & 2   & 4   &  4   & 0 & -1.281840         \\       
  2   & 4  &  1  &  5  &   1  &  0  &   5.491716          &            2  &  5   & 1   & 5   &  4   & 0 &  1.246601         \\       
  2   & 4  &  3  &  6  &   1  &  0  &   2.778464          &            2  &  4   & 2   & 4   &  4   & 0 & -5.746248         \\       
  2   & 4  &  3  &  4  &   1  &  0  &  -1.876604          &            2  &  4   & 1   & 5   &  4  & 0 & -3.402626         \\      
  2   & 4  &  1  &  6  &   1  &  0  &   1.515858          &            1  &  5   & 1   & 5   &  4   & 0 & -5.378448         \\       
  2   & 4  &  1  &  4  &   1  &  0  &   0.373965          &            2  &  5   & 2   & 5   &  5 &   0 & -4.184793         \\       
    \end{tabular}                                                                                                                          
  \end{ruledtabular}
  \caption{CCEI two-body matrix-elements for the $sd$-shell.}
  \label{tab2}
\end{table*}